\documentstyle[seceq,letter,psfig]{ptptex}

\newcommand{\beq}{\begin{equation}}
\newcommand{\eeq}{\end{equation}}
\newcommand{\beqa}{\begin{eqnarray}}
\newcommand{\eeqa}{\end{eqnarray}}

\newcommand{\simg}
   {\mathrel{\raise.3ex\hbox{$>$\kern-.75em\lower1ex\hbox{$\sim$}}}}
\newcommand{\siml}
   {\mathrel{\raise.3ex\hbox{$<$\kern-.75em\lower1ex\hbox{$\sim$}}}}

\preprintnumber[3cm]{
KUNS-1744\\ YITP-01-75\\ hep-ph/0111086}

\title{
Quintessence Cosmology and Varying $\alpha$
}

\author{
Takeshi Chiba~$^a$ and Kazunori Kohri~$^b$
}

\inst{
$^a$~Department of Physics, Kyoto University,
Kyoto 606-8502, Japan \\
$^b$~Yukawa Institute for Theoretical Physics, Kyoto University, 
Kyoto, 606-8502 Japan 
}

\recdate{
}

\abst{
If the reported measurements of the time variation of the fine structure 
constant from observations of distant QSOs are correct, 
combined with the Oklo limit they would strongly constrain 
the class of the quintessence potential. 
If these results prove valid, future satellite experiment (STEP) 
should measure the induced violation of the weak equivalence principle. 
Future cosmological observations of nearby $(z \siml 0.5)$ absorption 
systems would make it clear whether the variation is significant or not.
}

\begin{document}

\maketitle

{\bf 1. Introduction.} 
The Universe is filled with dark energy. If dark energy is dynamical,
its dynamics  is described by an ultra-light scalar field with 
mass $\siml H_0$, called ``quintessence''. Such a field could interact 
with ordinary matter, unless forbidden by symmetries. Because this field
would be dynamical and the exchange of light fields gives rise to long
range forces, the interaction of the field with ordinary matter would
result in the time variation of the constants of nature over
cosmological time scales and the violation of the weak equivalence
principle. Since those effects have not been observed
\cite{Chiba:2001ui}, such direct couplings were believed to be strongly 
suppressed.\cite{Carroll:1998zi,Chiba:1999wt}

Recently, however, observations of
a number of absorption systems in the spectra of distant quasars
indicate a smaller value of $\alpha$ in the past, and an optical sample 
exhibits a 4$\sigma$ deviation for $0.5 < z < 3.5$: 
$\Delta \alpha/\alpha = (-0.72\pm 0.18)\times 10^{-5}$.\cite{Webb:2001mn} 
On the other hand, the severest limit on $\dot \alpha$ obtained 
from analysis of the isotope abundances in the Oklo natural reactor 
operated 1.8 Gyr ago is $|\Delta \alpha|/\alpha \siml 10^{-7}$. 
Are these data compatible?

In this paper, we elucidate the discrepancy between the QSO data and 
the Oklo limit and then point out the potential significance of these 
data in constraining a model of quintessence. We also estimate the degree of 
the violation of the equivalence principle to motivate future
experimental precision tests of the equivalence principle.\cite{Dvali:2001dd}

While we were preparing this paper for the submission, we became aware of  
two related papers.\cite{Dvali:2001dd,Olive:2001vz} 
Reference \cite{Dvali:2001dd} investigates a fifth force-type long range 
interaction mediated by a scalar field. Reference \cite{Olive:2001vz} 
studies a model with a large coupling
between non-baryonic dark matter and a scalar field.

{\bf 2. QSO, Oklo, and Quintessence. }
{}From the perspective of an effective theory, no couplings of the 
quintessence field to ordinary matter should be ignored, unless 
they are forbidden by symmetry.\cite{Carroll:1998zi} Therefore, we expect, 
for example, the coupling between $\phi$ and the photon\footnote{We note that 
due to the approximate global symmetry 
$\phi\rightarrow \phi +{\rm const.}$, the only possible coupling of 
quintessence axion to a photon is of the form, $\phi F\widetilde{F}$,  
and hence a $\phi FF$-type coupling is absent \cite{Kaplan:1985dv}. 
Therefore the QSO data cannot be explained by quintessence axion.} 
\beq
f(\phi)F_{\mu\nu}F^{\mu\nu},
\label{couple}
\eeq
where $f(\phi)$ is a function of $\phi$. Therefore the fine structure 
constant becomes a function of $\phi$: $\alpha=\alpha(\phi)$. 
We may expand $\alpha(\phi)$ about the present value $\phi_{\rm now}$ 
assuming $\phi-\phi_{\rm now} < M_{pl}$:\footnote{We note, 
however, that generically $\phi \sim M_{pl}$ for quintessence since 
$V''\simeq H_0^2$. Hence this assumption is barely valid.} 
\beq
\alpha(\phi)=\alpha_{\rm now}+\lambda 
\left({\phi-\phi_{\rm now}\over M_{pl}}\right) + 
\dots.
\label{alpha}
\eeq

\begin{figure}[htdp]
  \begin{center}
  \leavevmode\psfig{figure=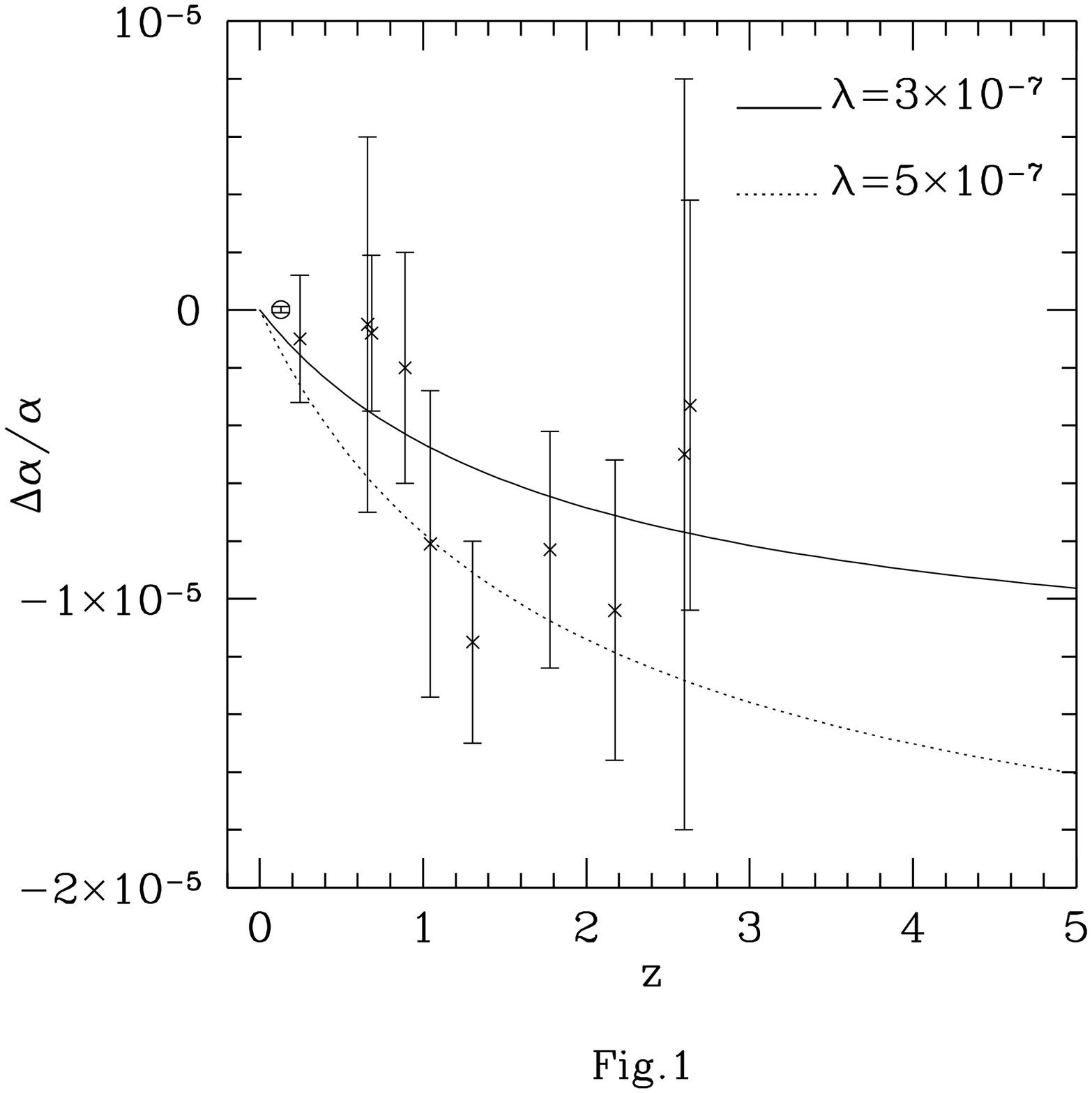,width=8cm}
  \end{center}
  \caption{
 $\Delta\alpha/\alpha$ as a function of $z$. The points marked by the
 symbols ``X'' are the QSO data of Webb et al. 
 and the open circle at $z=0.13$ is the Oklo datum. 
 The form of $\Delta\alpha/\alpha$ for 
 quintessence with an inverse power-law potential $\propto \phi^{-2}$
 and with $\Omega_M=0.3$ is shown for $\lambda=3\times 10^{-7}$ (solid
 curve) and for $\lambda=5\times 10^{-7}$ (dashed curve).
}
\label{fig:fig1}
\end{figure}

If no symmetry is imposed, the variation of $\alpha$ with $\phi$ is 
generally written to the leading order in $\phi/M_{pl}$ as
\beq
\label{eq:delaa}
{\Delta\alpha\over \alpha}\simeq {\lambda\Delta\phi\over \alpha_0 M_{pl}},
\eeq 
where $\Delta\phi\equiv \phi_{\rm then}-\phi_{\rm now}$, and
$\phi_{\rm then}$ denotes the value of $\phi$ at that time.
Observational evidence indicates 
$\Delta\alpha/\alpha \simeq -10^{-5}$ for $0.5 < z < 3.5$,
 which in turn implies 
\beq
\lambda \left({\Delta\phi\over M_{pl}}\right) \simeq -10^{-7}.
\label{limit:qso}
\eeq
Such time variation can be explained for a wide class of quintessential 
potentials. An example is shown in Fig. 1 for the inverse power-law 
potential $\propto \phi^{-2}$ with $\Omega_M=0.3$ and $h=0.65$. 

We also plot the datum from the Oklo phenomenon. 
The Oklo natural reactor that operated about 1.8 
billion years ago in Oklo, Gabon corresponding to 
$z\simeq 0.13$\footnote{The redshift 
depends mainly on the Hubble parameter. We assume $h=0.65$. However, 
its exact value is not important to our argument. What is important is 
$z_{\rm Oklo}\siml 0.5$.} yields a bound of 
 $\Delta\phi/\phi=(-0.9\sim 1.2)\times 10^{-7}$ or 
$(-6.7\sim 5.0)\times 10^{-17} {\rm yr}^{-1}$ \cite{Damour:1996zw}. 
Using new samples that were carefully collected to minimize natural 
contamination with a careful temperature estimate of the reactors, 
Fujii et al. derived the bound $(-0.36\sim 1.44)\times 10^{-8}$ or 
$(-0.2\pm 0.8)\times 10^{-17} {\rm yr}^{-1}$ \cite{Fujii:2000kn}.
These bounds imply 
\beq
\lambda \left({|\Delta\phi |\over M_{pl}}\right) \siml 10^{-9} \sim 10^{-10}.
\label{limit:oklo}
\eeq 
In Fig.~2 the required changes of $\phi$ are exhibited as functions
of $z$, where we have adopted the conservative bound of Damour and
Dyson. Figure 2 shows that the (absolute) value of $\phi$ must have 
decreased by more than two orders of magnitude only recently, 
$z\siml 1$. It is noted, however, that the Oklo bound is not 
cosmological but geophysical in nature, and assigning to the Oklo event 
the ``cosmological redshift'' $z\simeq 0.13$ may not be 
justified.\cite{barrow} For this reason, other {\it cosmological} 
observations of nearby absorption systems ($z \siml 0.5$) are required
to verify or contradict such a conclusion. 
The current result is consistent with a null result\cite{Carilli:2000ca}.

The two relations in Eqs.(\ref{limit:qso}) and (\ref{limit:oklo}) imply 
that the scalar field stopped evolving abruptly after $z\simeq 1$. 
Therefore, if we interpret the recent QSO data {\it and} the Oklo
naively and attempt to explain them in terms of quintessence 
coupling to a photon, then we are led to a model in which a potential
has a local minimum into which the scalar field was trapped only 
recently ($z \siml 1$), which would require a fine-tuning of model 
parameters, or to a model with a large coupling between non-baryonic 
dark matter and the scalar field.\cite{Olive:2001vz}
Examples are the Albrecht-Skordis model \cite{Albrecht:2000rm},  
the Barreiro-Copeland-Nunes model \cite{Barreiro:2000zs} and 
supergravity-inspired models,\cite{Copeland:2000vh,Brax:2001ah} to name a few.

\begin{figure}[htdp]
\centerline{\psfig{figure=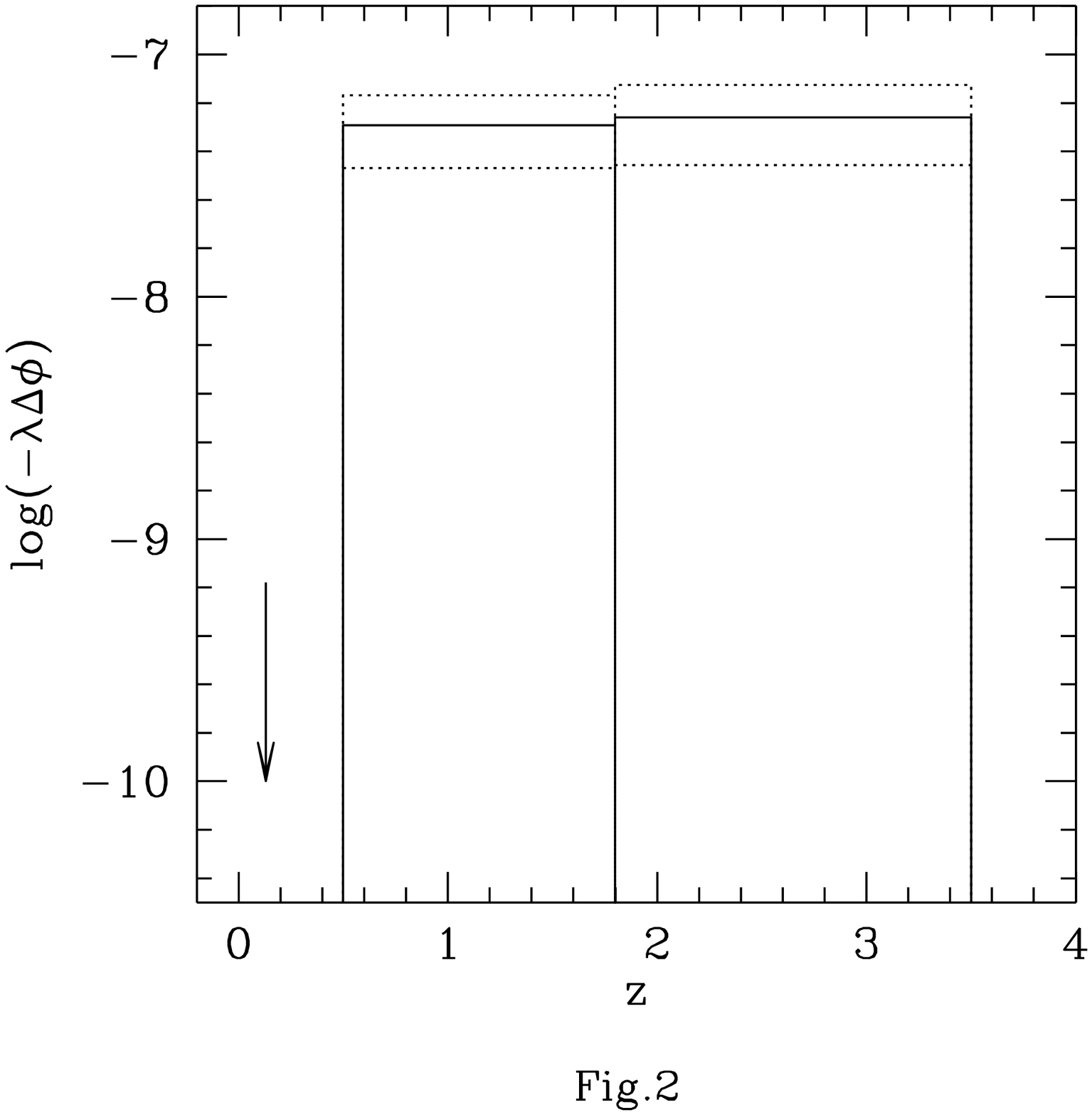,width=8cm}}
  \caption{
$\lambda\Delta\phi$ as a function of $z$. The histogram represents 
the QSO data of Webb et al. 
Errors are represented by dotted lines. 
The arrow indicates the upper limit from the Oklo data. 
$\Delta\phi=\phi_{\rm then}-\phi_{\rm now}$. This figure shows
clearly that $\phi$ must have stopped evolving abruptly only 
recently, $z < 1$.
}
\label{fig:fig2}
\end{figure}

{\bf Equivalence Principle. }
The direct coupling of the form in Eq.(\ref{couple}) induces a 
violation of the weak coupling principle because the baryon mass is
then a function of $\phi$. The degree of violation of the equivalence 
principle may be estimated in the manner of Ref.\cite{Dvali:2001dd}. 
(The details of this calculation appear in the Appendix.) 
Changing $\alpha$ causes a change in the nucleon mass coming 
from electromagnatic radiative corrections and hence results in a  
composite-dependence in free-fall experiments. The conventional 
E\"{o}tv\"{o}s parameter $\eta$, which measures the difference 
between the accelerations of two test bodies, is estimated to be
\begin{eqnarray}
    \label{eq:eta_present}
    \eta \sim 10^{-17}\left({\lambda\over 3\times 10^{-7}}\right)^2.
\end{eqnarray}
This value is much smaller than the present upper bound given by 
E\"{o}tv\"{o}s-Dicke-Braginsky type experiments, $\eta <
10^{-13}$,\cite{baessler:1999} and is therefore consistent with this
bound. We note, however, that proposed satellite experiments, such as
STEP (Satellite Test of the Equivalence Principle) \cite{step}, should be 
able to detect such a violation. STEP should be sensitive to a 
violation in the range $\eta \sim 10^{-13}- 10^{-18}$.

{\bf 3. Summary. }
Assuming that dark energy is an ultra-light scalar field 
(or ``quintessence''), we have obtained quantitative results for 
the required cosmological change of the scalar field 
to account for both the recent QSOs data and the Oklo datum. 
If the reported measurements of nonzero $\Delta\alpha/\alpha$ are
corect, combined with the Oklo limit, they would strongly constrain the 
class of quintessence potential. 
If these QSOs observations prove valid, the proposed satellite
experiment for testing the equivalence principle (STEP) should be able
to detect the violation of the weak 
equivalence principle induced by a scalar force mediated by quintessence.
Because of the geophysical nature of the Oklo bound, its validity with
respect to the present problem is suspect, and it is hoped that 
future cosmological observations of nearby $(z\siml 0.5)$ absorption 
systems may clarify the situation. 

{\bf Noted added:} The great difficulty of explaining the
observed time variation of the fine structure constant from the
viewpoint of particle physics is discussed in the recent work
Ref.\cite{bdd}: the induced variation in the vacuum energy would be 
enormously large. However, this may be nothing but the aspect of the 
problem of the cosmological constant: how the vacuum energy gravitates. 

\section*{Acknowledgements}
T. C. was supported in part by a Grant-in-Aid for Scientific Research 
(No. 13740154) from the Japan Society for the Promotion of Science.

\appendix
\section{}

In this appendix, we calculate the degree of violation of the weak 
equivalnce principle induced by varying $\alpha$ following the approach
of Ref.\cite{Dvali:2001dd}. 

The modification of the nucleon mass results from the electromagnetic
radiative corrections. The leading order of the correction of proton
and neutron masses is represented by~\cite{Gasser:1982ap}
\begin{eqnarray}
    \label{eq:mass_corr}
    \delta m_i = \delta \tilde{m}_i \frac{\Delta \alpha}{\alpha},
\end{eqnarray}
where ``$i$'' is $p$ for the proton and $n$ for neutron, and 
\begin{eqnarray}
    \label{eq:tildeB}
    \delta \tilde{m}_p &\simeq& 0.63 \ {\rm MeV}, \\
    \delta \tilde{m}_n &\simeq& -0.13 \ {\rm MeV}.
\end{eqnarray}
In this situation, the nucleon mass $m_i$ is not constant, but
depends on $\phi$. Then, the nucleon-$\phi$ coupling induces the
effective Yukawa interaction
\begin{eqnarray}
    \label{eq:yukawa}
    {\cal L}_{\rm int} = m_i (\phi) \overline{N_i} N_i = g_i \phi
    \overline{N_i} N_i,
\end{eqnarray}
where $N_i$ represents the spinor of nucleon ``$i$''. Using
Eq.~(\ref{eq:delaa}), the coupling constant $g_i$ is given by
\begin{eqnarray}
    \label{eq:gN_phi}
    g_i = \frac{\delta \tilde{m}_i \lambda}{\alpha_0 M_{pl}}.
\end{eqnarray}
Therefore, the exchange of $\phi$ induces an extra-ordinary
scattering among nucleons and leads to the Yukawa potential
\begin{eqnarray}
    \label{eq:yukawa_pot}
    V(r) = - \sum_i \sum_j \frac{g_i g_j}{4 \pi} \frac{e^{- m_{\phi}
    r}}{r} n^E_{i} n_j,
\end{eqnarray}
where $r$ is the distance between the Earth and the test body,
$m_{\phi}$ is the mass of $\phi$, and $n^E_{i}$ ($n_j$) is the
number of the nucleons in the Earth (test body). These couplings
depend on the nucleon species, which leads to the violation of the
equivalence principle. Because the mass of $\phi$ is minuscule ($m_{\phi}
\sim H_0$), we limit our consideration to the following shape of 
the potential in this study:
\begin{eqnarray}
    \label{eq::yukawa_pot2}
    V(r) = - \sum_i \sum_j \frac{g_i g_j}{4 \pi} \frac{1}{r} n^E_{i}
    n_j, \quad {\rm for} \ r \ll \frac{1}{m_{\phi}} \sim H_o^{-1}.
\end{eqnarray}

The acceleration induced by the $\phi$-exchange force is given by
\begin{eqnarray}
\label{eq:a_phi}
    a_{\phi} = \frac{1}{m} \frac{dV(r)}{dr},
\end{eqnarray}
where $m$ is the mass of the test body.  On the other
hand, the usual Newtonian acceleration is given by
\begin{eqnarray}
    \label{eq:a_g}
    a_g = \frac{M_{E}}{M_{pl}^2 r^2}, 
\end{eqnarray}
with the mass of the Earth $M_{E}$. Then, the total acceleration is
given by
\begin{eqnarray}
    \label{eq:a_tot}
    a = a_{\phi} + a_g.
\end{eqnarray}
It is convenient to introduce the following parameter to study
the difference between the accelerations of two test bodies in
E\"{o}tv\"{o}s-Dicke-Braginsky-type experiment:~\cite{Eotvos:1922pb}
\begin{eqnarray}
    \label{eq:eta}
    \eta = 2\frac{|a_1 - a_2|}{|a_1 + a_2|} .
\end{eqnarray}
Here $a_1$ and $a_2$ are the accelerations of the two bodies. We
assume that the test bodies have almost equal masses, $m_1
\simeq m_2$, i.e., $n_{n,1} + n_{p,1} \simeq n_{n,2} + n_{p,2}$. In
addition, we assume that $m = (n_n + n_p)\overline{m}$ and $M_E
= (n^E_n + n^E_p)\overline{m} $, where $\overline{m}$ denotes the
atomic mass unit ($\simeq 0.931$ MeV). Then, for relatively small
$\lambda$ ($\ll {\cal O}(10)$), we find
\begin{eqnarray}
    \label{eq:eta_eval}
    \eta &\simeq& \frac{|a_{\phi, 1} - a_{\phi, 2}|}{a_g} \nonumber \\
    &\simeq& \frac{\lambda^2}{4\pi \alpha_0^2} \frac1{ \overline{m}^2}
    \left(R^E_n\delta \tilde{m}_n+R^E_p\delta \tilde{m}_p\right)
    \left(\Delta R_n\delta \tilde{m}_n+\Delta R_p\delta \tilde{m}_p\right),
\end{eqnarray}
where we have defined the nucleon-number fraction in the Earth as 
$R^E_i \equiv n^E_i/(n^E_n+n^E_p)$ and the difference of the
nucleon-number fraction of the test bodies as 
$\Delta R_i \equiv |n_{i,1} - n_{i,2}|/(n_n+n_p)$. For the sake of 
simplicity, we assume that $R^E_n
\simeq R^E_p \simeq 0.5$. In Ref.~\cite{Dvali:2001dd} the relation 
$\Delta R_n \sim \Delta R_p \sim 0.06 - 0.1$ is estimated for typical
materials used in experiments (copper, lead and
uranium). Adopting the above values, we find
\begin{eqnarray}
    \label{eq:eta_eval2}
    \eta \sim 10^{-4} \lambda^2.
\end{eqnarray}



\begin{thebibliography}{99}

\bibitem{Chiba:2001ui}
Recent experimental constraints on $\dot\alpha$ and $\dot G$ are reviewed in, 
T.~Chiba,
arXiv:gr-qc/0110118, in the proceedings of {\it Frontier of Cosmology 
and Gravitation}.

\bibitem{Carroll:1998zi}
S.~M.~Carroll,
Phys.\ Rev.\ Lett.\  {\bf 81}, 3067 (1998).

\bibitem{Chiba:1999wt}
T.~Chiba,
Phys.\ Rev.\ D {\bf 60}, 083508 (1999).

\bibitem{Webb:2001mn}
J.~K.~Webb {\it et al.},
Phys.\ Rev.\ Lett.\  {\bf 87}, 091301 (2001).

\bibitem{Dvali:2001dd}
G.~R.~Dvali and M.~Zaldarriaga,
arXiv:hep-ph/0108217.

\bibitem{Olive:2001vz}
K.~A.~Olive and M.~Pospelov,
arXiv:hep-ph/0110377.

\bibitem{Kaplan:1985dv}
D.~B.~Kaplan,
Nucl.\ Phys.\ B {\bf 260}, 215 (1985); 
M.~Srednicki,
Nucl.\ Phys.\ B {\bf 260}, 689 (1985).

\bibitem{Damour:1996zw}
T.~Damour and F.~Dyson,
Nucl.\ Phys.\ B {\bf 480}, 37 (1996).

\bibitem{Fujii:2000kn}
Y. Fujii, A. Iwamoto, T. Fukahori, T. Ohnuki, M. Nakagawa, H. Hikida, 
Y. Oura and P. M\"{o}ller,
Nucl.\ Phys.\ B {\bf 573}, 377 (2000).

\bibitem{barrow}
J.D. Barrow and C. O'Toole, MNRAS {\bf 322}, 585 (2001).

\bibitem{Carilli:2000ca}
C.~L.~Carilli {\it et al.},
Phys.\ Rev.\ Lett.\  {\bf 85}, 5511 (2000).

\bibitem{Albrecht:2000rm}
A.~Albrecht and C.~Skordis,
Phys.\ Rev.\ Lett.\  {\bf 84}, 2076 (2000).

\bibitem{Barreiro:2000zs}
T.~Barreiro, E.~J.~Copeland and N.~J.~Nunes,
Phys.\ Rev.\ D {\bf 61}, 127301 (2000).

\bibitem{Copeland:2000vh}
E.~J.~Copeland, N.~J.~Nunes and F.~Rosati,
Phys.\ Rev.\ D {\bf 62}, 123503 (2000).

\bibitem{Brax:2001ah}
P.~Brax, J.~Martin and A.~Riazuelo,
Phys.\ Rev.\ D {\bf 64}, 083505 (2001).

\bibitem{baessler:1999}
S. Baessler et al., Phys. Rev. Lett. {\bf 83}, 3585 (1999).

\bibitem{step}
http://einstein.stanford.edu/STEP/

\bibitem{Gasser:1982ap}  
J.~Gasser and H.~Leutwyler,
Phys.\ Rept.\  {\bf 87}, 77 (1982).

\bibitem{Eotvos:1922pb}
R.~V.~E\"{o}tv\"{o}s, D.~Pekar and E.~Fekete,
Annalen Phys.\  {\bf 68}, 11 (1922); 
P.~G.~Roll, R.~Krotkov and R.~H.~Dicke,
Annals Phys.\  {\bf 26}, 442 (1964); 
V.B. Braginsky and V.I. Panov,
Sov. Phys. JETP {\bf 34}, 463 (1972).

\bibitem{bdd}
T. Banks, M. Dine, and M.R. Douglas, hep-ph/0112059.

\end{thebibliography}
\end{document}